# INVERSE CHAOS SYNCHRONIZATION BETWEEN THE UNI-DIRECTIONALLY COUPLED SYSTEMS WITH MODULATED MULTIPLE TIME DELAYS

E. M. Shahverdiev

Institute of Physics, H.Javid,33,Baku,AZ1143,Azerbaijan

## ABSTRACT

We present the first report on inverse chaos synchronization where a driven modulated multiple time-delay chaotic system synchronizes to the inverse state of the driver system. Numerical simulations fully support the analytical approach.

PACS number(s):05.45.Xt, 05.45.Vx, 42.55.Px, 42.65.Sf

Key words: variable time delays, inverse chaos synchronization, Ikeda model.

## I.INTRODUCTION

Chaos synchronization [1] as a chaos control method is of fundamental importance in a variety of complex physical, chemical and biological systems [2-3]. Synchronization of coupled chaotic systems eliminates some degrees of freedom of the coupled system and so produces a significant reduction of complexity. The occurrence of synchronization between elements of a large system allows significant simplification of computational and theoretical analysis of the system.

Synchronization phenomena in coupled systems have been especially extensively studied in the context of laser dynamics, electronic circuits, chemical and biological systems [2-3]. Application of chaos control can be found in secure communication, optimization of nonlinear system performance, modeling brain activity, species population control, see e.g.[2-3] and references therein.

Time-delayed systems are ubiquitous in nature, technology and society because of finite signal transmission times, switching speeds and memory effects [4]. Therefore the study of chaos synchronization in these systems is of considerable practical significance. Because of their ability to generate high-dimensional chaos, time-delay systems are good candidates for secure communications based on chaos synchronization [3].
Recently it has been shown [5] that the incorporation of additional time delays can play a stabilizing role in chaos synchronization and also can offer a higher complexity of dynamics than achievable in more conventional single delay time systems.

Identification of time delays of such systems can allow the eavesdropper to extract the message successfully using a simple local reconstruction of the time-delayed system. Quite recently it is established that variable time delay systems can offer an enhanced robustness to eavesdropper attack compared to the case of fixed time-delayed systems [6]. The fundamental requirement for enabling message extraction in a chaos-based communication is the ability to achieve high-quality synchronization [3]. In addition, in comparison with fixed time-delays, systems with variable (whether delibrate or unwanted) multiple time-delays are more realistic models in the interacting complex systems.

This paper presents the first report of the inverse chaos synchronization between uni-directionally coupled modulated multiple time delayed Ikeda models with two feedbacks. (For inverse synchronization [7], a time-delayed chaotic system $x$ drives another system $y$ in such a way that a driven system synchronize to the inverse state of the driver system: $x(t) = -y(t)$.)

## II. SYSTEM MODEL

Consider inverse synchronization between variable time-delayed Ikeda systems with two feedbacks,

$$\frac{dx}{dt} = -\alpha x - m_1 \sin x_{\tau_1} - m_2 \sin x_{\tau_2} \tag{1}$$

$$\frac{dy}{dt} = -\alpha y - m_3 \sin y_{\tau_1} - m_4 \sin y_{\tau_2} + K \sin x_{\tau_3} \tag{2}$$

This investigation is of considerable practical importance, as the equations of the class B lasers with feedback (typical representatives of class B are solid-state, semiconductor, and low pressure $CO_2$ lasers [8-9]) can be reduced to an equation of the Ikeda type [10].
The Ikeda model was introduced to describe the dynamics of an optical bistable resonator, playing an important role in electronics and physiological studies and is well-known for delay-induced chaotic behavior [10,11]. Physically $x$ is the phase lag of the electric field across the resonator; $\alpha$ is the relaxation coefficient for the driving $x$ and driven $y$ dynamical variables; $\tau_{1,2} = \tau_{01,02} + x_1(t)\tau_{a1,a2}\sin(\omega_{1,2}t)$ are the variable feedback loop delay times; $\tau_3 = \tau_{03} + x_1(t)\tau_{a3}\sin(\omega_3 t)$ is the variable time of flight between master laser $x$ and slave laser $y$; $\tau_{01,02,03}$ are the zero-frequency component, $\tau_{a1,a2,a3}$ are the amplitude, $\frac{\omega_{1,2,3}}{2\pi}$ are the frequency of the modulations; $x_1(t)$ is the output of system (1) for constant time delays, i.e. $\tau_1 = \tau_{01}, \tau_2 = \tau_{02}$; $m_{1,2}$ and $m_{3,4}$ are the feedback strengths for the master and slave systems, respectively; $K$ is the coupling strength between the systems.
Before considering the case of modulated time delays, we present the conditions for inverse synchronization for fixed time delays, i.e. $\omega_{1,2,3} = 0$.
We find that systems (1) and (2) can be synchronized on the inverse synchronization regime

$$x = -y \tag{3}$$

as the synchronization error signal $\Delta = x - (-y) = x + y$ for small $\Delta$ under the condition

$$\tau_{01} = \tau_{03}, m_1 + m_3 = K, m_2 = m_4 \tag{4}$$

obeys the following dynamics

$$\frac{d\Delta}{dt} = -\alpha\Delta - m_3 \Delta_{\tau_{01}} \cos x_{\tau_{01}} - m_4 \Delta_{\tau_{02}} \cos x_{\tau_{02}} \tag{5}$$

It is obvious that $\Delta = 0$ is a solution of system (5).

We study the stability of the synchronization regime $x = -y$ by using the Krasovskii-Lyapunov functional approach. According to [4], the sufficient stability condition for the trivial solution $\Delta = 0$ of time-delayed equation $\frac{d\Delta}{dt} = -r(t)\Delta + s_1(t)\Delta_{\tau_1} + s_2(t)\Delta_{\tau_2}$ is: $r(t) > |s_1(t)| + |s_2(t)|$. Therefore by using the Razumikhin-Lyapunov functional approach we obtain that the sufficient stability condition for the synchronization manifold $x = -y$ can be written as:

$$\alpha > |m_3| + |m_4| \qquad (6)$$

As Eq.(5) is valid for small $\Delta$ stability condition (6) found above, holds locally. Conditions (4) are the existence conditions for the synchronization manifold (3) between unidirectionally coupled Ikeda systems (1) and (2) with multiple delays.

Analogously by investigating corresponding error dynamics we also find that $x = -y$ is the inverse synchronization regime between systems (1) and (2) with the existence $\tau_{02} = \tau_{03}, m_2 + m_4 = K$ and $m_1 = m_3$ and stability conditions $\alpha > |m_3| + |m_4|$.

In the case of variable time delays establishing the existence and stability conditions for the synchronization is not as straightforward as for the constant time delays. Having in mind that for $\omega = 0$ we obtain a case of constant time delays, then as an initial guess one can benefit from the existence conditions for the constant time delays case. It is our conjecture that high quality synchronization $x = -y$ will be obtained if the parameters satisfy the conditions: $m_1 + m_3 = K, m_2 = m_4$ if $\tau_1(t) = \tau_3(t)$ (i.e. the zero frequency components, amplitudes, and modulation frequencies are the same: $\tau_{01} = \tau_{03}, \tau_{a1} = \tau_{a3}, \omega_1 = \omega_3$), or $m_2 + m_4 = K, m_1 = m_3$ if $\tau_2(t) = \tau_3(t)$ (i.e. $\tau_{02} = \tau_{03}, \tau_{a2} = \tau_{a3}, \omega_2 = \omega_3$). As evidenced by the numerical simulations below, this conjecture is found to be well-based.

## III. NUMERICAL SIMULATIONS

We studying the synchronization between the laser systems with variable time delays using the cross-correlation coefficient $C$ [12]

$$C(\Delta t) = \frac{<(x(t)-<x>)(y(t+\Delta t)-<y>)>}{\sqrt{<(x(t)-<x>)^2><(y(t+\Delta t)-<y>)^2>}}, \qquad (7)$$

for $x = y$: where $x$ and $y$ are the outputs of the interacting laser systems; the brackets $<.>$ represent the time average; $\Delta t$ is a time shift between laser outputs. This coefficient indicates the quality of synchronization: $C = \pm 1$ means perfect (inverse) synchronization.

As mentioned above, in chaos based communication schemes synchronization between the transmitter and receiver systems are vital for message decoding. With this in mind we present here the first report of inverse chaos synchronization between variable time-delayed Ikeda models.

Figure 1 portrays time series of the master system (x-solid line) and slave system (y-dotted line) for inverse chaos synchronization $x = -y$ between unidirectionally coupled lasers, Eqs.(1) and (2) for variable feedback time delays $\tau_1(t) = 4 + 3\sin(0.1t), \tau_2(t) = 7 + 3\sin(0.1t)$ and variable coupling time delay $\tau_3(t) = 4 + 3\sin(0.1t)$ with parameter values as $\alpha = 4, m_1 = 7.5, m_2 = m_4 = 2.5, m_3 = 1.4, K = 8.9$.

Figure 2 depicts synchronization error dynamics $\Delta = x + y$ versus time for parameters as in figure 1. C =1 is the cross-correlation coefficient between the master (transmitter) and slave (receiver) system outputs.

The value of the cross-correlation coefficient for the case testifies to the high quality chaos synchronization, which is vital for information processing in chaos-based communication systems.

## IV. CONCLUSIONS

To summarize, we have reported on inverse chaos synchronization in unidirectionally coupled variable multiple time delayed Ikeda systems. The results are of certain importance for information processing in chaos based communication systems.

## V. ACKNOWLEDGEMENTS

This research was supported by a Marie Curie Action within the $6^{th}$ European Community Framework Programme Contract N:MIF2-CT-2007-039927-980065.

Figure captions

FIG.1. Numerical simulation of unidirectionally coupled variable time delay systems, Eqs.(1-2) for $\alpha = 4, m_1 = 7.5, m_2 = 2.5, m_3 = 1.4, m_4 = 2.5, K = 8.9$ and $\tau_1(t) = 4 + 3x_1(t)\sin(0.1t), \tau_2(t) = 7 + 3x_1(t)\sin(0.1t), \tau_3(t) = 4 + 3x_1(t)\sin(0.1t)$. Inverse synchronization: Time series of the master system (x-solid line) and slave systems (y-dotted line). Dimensionless units.

FIG.2. Numerical simulation of unidirectionally coupled variable time delay systems, Eqs.(1-2). Error dynamics, $\Delta = x + y$ versus time. The parameters are as in figure 1. C is the cross-correlation coefficient between the master and slave systems. Dimensionless units.

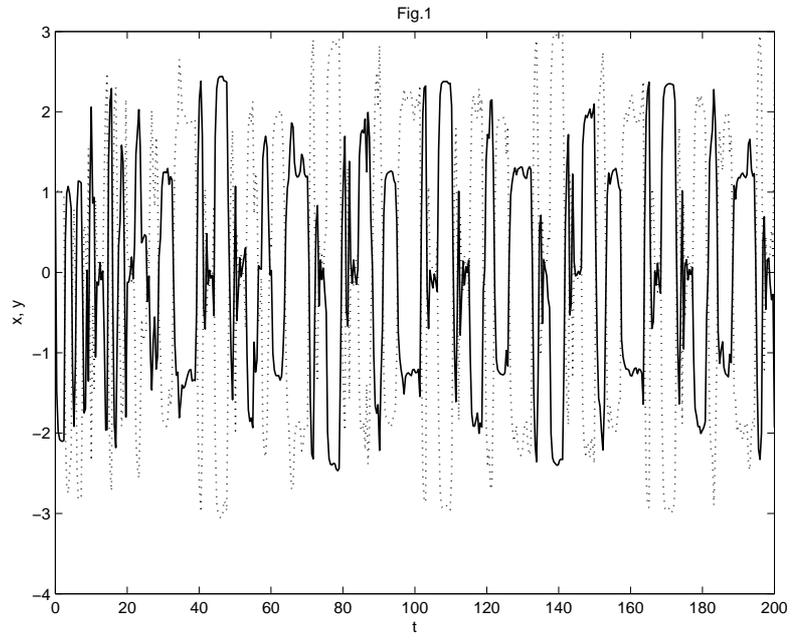

Fig.1

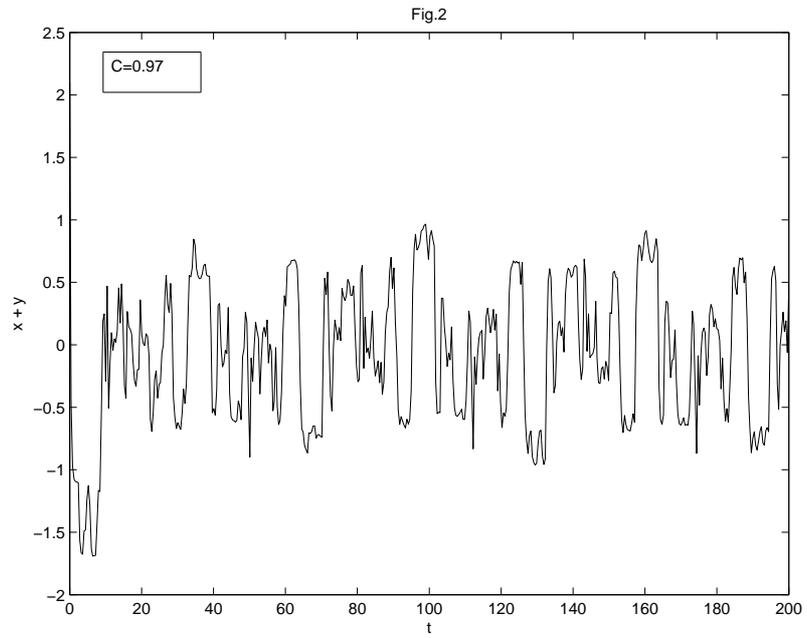

Fig.2